\documentclass[12pt,letterpaper]{article}
\usepackage[plain]{fullpage}

\usepackage{latexsym}

\newtheorem{theorem}{Theorem}
\newtheorem{corollary}[theorem]{Corollary}

\def\F{{\mathbf F}}

\newcommand{\ket}[1]{{|{#1}\rangle}}

\newcommand{\size}[1]{| #1 |}

\newcommand{\poly}{\mathrm{poly}}

\newcommand\Mat{{\mathcal{M}}}
\newcommand\GL{{\mathrm{GL}}}
\newcommand\PGL{{\mathrm{PGL}}}
\newcommand\SL{{\mathrm{SL}}}
\newcommand\PSL{{\mathrm{PSL}}}
\newcommand\Tr{{\mathrm{Tr}}}
\newcommand\tr{{\mathrm{tr}}}

\begin{document}


\title{Finding hidden Borel subgroups of the general linear group}

\author{
G\'abor Ivanyos
\\
Computer and Automation Research Institute
\\
of the Hungarian Academy of Sciences, 
\\
Kende u. 13-17,
H-1111, Budapest, Hungary
\\
E-mail:
{\tt Gabor.Ivanyos@sztaki.hu}
}

\maketitle

\begin{abstract}
We present a quantum algorithm for solving
the hidden subgroup problem
in the general linear group over a finite field
where the hidden subgroup is promised to
be a conjugate of the group of the invertible
lower triangular matrices. The complexity of
the algorithm is polynomial when size of the base field
is not much smaller than the degree.
\end{abstract}


\section{Introduction}
\label{sect:intro}

The hidden subgroup problem (HSP for short) 
is the following. We are given a black box function $f$ on a group $\cal G$  
such that there is a subgroup $\cal H$ of $\cal G$ satisfying
$f(x)=f(y)$ if and only if $x$ and $y$ are in
the same right coset of $\cal H$ (that is, $yx^{-1}\in {\cal H}$).
The task is to determine
the subgroup $\cal H$, which is unique and
called the subgroup hidden by $f$.  Computing orders of elements of
groups, calculating discrete logarithms and even
finding isomorphisms between graphs can be cast
in the paradigm of the HSP~\cite{jozsa01}.

On classical computers, the query complexity of the hidden
subgroup problem is exponential (in $\log\size{\cal G}$) 
already over finite commutative groups. 
In the quantum setting $f$ is assumed to be given by a quantum oracle
which is a unitary map $U_f$ mapping states of
the form $\ket{x}\ket{0}$ to $\ket{x}\ket{f(x)}$.
In contrast to the classical case the quantum
query complexity of the HSP
is polynomial (in $\log\size{\cal G}$), see \cite{ehk04}.
Furthermore, there are polynomial time quantum
algorithms~\cite{bl95,kit95} solving the hidden subgroup problem
in abelian groups, generalizing 
Shor's result on order finding and computing discrete logarithm~\cite{Shor}.
Computing the structure of finite commutative black box 
groups~\cite{cm01} is a more general application of the
abelian HSP.

As the graph isomorphism problem is involved, 
in the past decade considerable efforts have been spent on 
finding efficient algorithms for noncommutative cases 
of the HSP. Although nice results have been achieved in this direction,
the groups in which the HSP can be solved at present
in quantum polynomial time remain actually very close to being commutative. 
One of the widest classes of finite groups in which the HSP is known to have a 
polynomial time quantum solution consists of solvable groups 
whose derived subgroup have of constant derived length and constant
exponent~\cite{FIMSS-stoc}. Other classes of groups with 
efficient quantum HSP algorithms include
certain "almost Hamiltonian" groups \cite{gavinsky04} and 
two-step nilpotent groups \cite{iss08}. The latter class
contains Heisenberg groups for which efficient
HSP algorithms are also given in \cite{bcvd05} and \cite{bacon08}. 
The "pretty good measurement" technique of \cite{bcvd05} actually 
works also in certain nilpotent semidirect product groups of higher 
nilpotency class. An efficient HSP algorithm is given 
in \cite{ilg07} for a special family of groups 
which possess a large commutative subgroup and 
a map transforming the HSP of the whole group to
the HSP of the abelian subgroup.
The limited success in finding good noncommutative HSP algorithms 
indicates that the problem may be actually difficult. 
The connection between the HSP in dihedral groups and 
some supposedly difficult lattice problem exposed 
in~\cite{Regev} provides further evidence for that. 

Putting restrictions on the class of the possible hidden subgroups
can result in efficient algorithms for finding them even in 
fairly noncommutative groups. Most importantly, the HSP for normal subgroups 
can be solved in quantum polynomial time in groups for which efficient 
quantum Fourier transforms exist (see~\cite{gsvv01} and~\cite{hrt03}) 
and in a class of groups including solvable groups and more \cite{ims03}.
The methods of \cite{mrrs04} and \cite{gpc09} work
efficiently for sufficiently large non-normal hidden
subgroups in certain semidirect products. 

The first polynomial time algorithm for finding special
hidden subgroups in simple and almost simple groups is given
in \cite{dmr10}. (An almost simple group 
has a large noncommutative simple constituent.)
The main result of Ibid.~is an efficient 
quantum algorithm that solves the HSP in the
group of $2$ by $2$ invertible matrices (and
related groups) where the hidden subgroup is
promised to be a so-called Borel subgroup (definition will be
given below). 
In this paper generalize this result to finding hidden 
Borel subgroups in general linear groups 
of higher degree. 

We denote by $\GL_n(\F_q)$
the general linear group consisting
of the invertible $n\times n$ matrices over the finite
field $\F_q$ having $q$ elements. We propose a
quantum algorithm for the HSP in $\GL_n(\F_q)$
where the hidden subgroup is promised to be 
a Borel subgroup. For brevity we
use the term {\em hidden Borel subgroup problem} for
this promise problem. Our algorithm works in polynomial
time if $q$ is not much smaller than $n$.

A Borel subgroup of $\GL_n(\F_q)$ is a conjugate
of the subgroup consisting of the invertible lower
triangular matrices (see~\cite{Springer}). An alternative definition for
a Borel subgroup is being the stabilizer in $GL_n(\F_q)$
of a flag $\F_q^n>U_1>U_2>\ldots>U_{n-1}>(0)$ of subspaces of
the space $V=\F^n$ of the column vectors of length $n$.
Indeed, for $0<k<n$ let $V_k$ be the
set of column vectors whose first $k$ entries
are zero. Then the 
invertible lower triangular matrices $A$ form
the stabilizer of the flag $\F_q^n>V_1>V_2>\ldots>V_{n-1}>(0)$,
and their conjugates $X^{-1}AX$ by $X$ form the stabilizer
of the flag $\F_q^n>X^{-1}V_1>X^{-1}V_2>\ldots>X^{-1}V_{n-1}>(0)$.
To fix a nicely defined output, by solving the hidden Borel subgroup 
problem we mean determining the flag of subspaces stabilized 
by the hidden Borel subgroup. We remark that, 
given such a flag, it is easy to construct
generators for its stabilizer.

Our method (described in Section~\ref{sect:Borel-GL}) 
is based on the observation that a coset of a 
Borel subgroup is quite a large subset
of a linear space of $n$ by $n$ matrices. The main
technical tool is a version of the standard algorithm for the 
abelian HSP, adapted to linear spaces (see Section~\ref{sect:QFT}). 
In Section~\ref{sect:Borel-SL} we show how to extend
our result to finding hidden Borel subgroups of the special
linear group.

\section{The quantum Fourier transform for linear spaces}

\label{sect:QFT}

In this section we briefly  overview the main
ingredient of the standard method for solving the 
hidden subgroup problem in $\F_q^m$ where the hidden 
subgroup is promised to be an $\F_q$-linear subspace 
$W$ of $\F_q^m$ and give an interpretation of
the result in the special case of a linear space of
matrices.

The procedure receives a superposition
\begin{equation}
\label{eq:cosetsup}
\frac{1}{\sqrt{\size {W}}}\sum_{v\in {W}}\ket{v+v_0}
\end{equation}
over a coset $W+v_0$ and obtains information on ${W}$ 
using the quantum Fourier transform (QFT) of the group 
$\F_q^m$. Here we use a version which is the $m$'th 
tensor power of the QFT defined 
in \cite{vdhi06} for $\F_q$. This transform
maps a $\ket{x}$ ($x\in \F_q$) to 
$$\frac{1}{\sqrt q}\sum_{y\in \F_q}\omega^{\Tr(xy)}\ket{y}$$
where $\Tr$ is the trace map from $\F_q$ to $\F_p$
and $\omega$ is the primitive $p$'th root of unity
$e^{\frac{2\pi i}{p}}$.
(Here $p$ is the prime such that $q=p^r$ for a
positive integer $r$
and the trace map is defined as 
$\Tr(x)=\sum_{i=0}^{r-1}x^{p^r}$.)
By Lemma~2.2 of \cite{vdhi06}, this map has a polynomial
time approximate implementation on a quantum computer,
therefore its $m$'th tensor power can be efficiently approximated
as well. The image of $\ket{x}$ for
a vector $x=(x_1,\ldots,x_m)^T\in\F_q^m$
under the tensor power map is
$$\frac{1}{q^{m/2}}\sum_{y\in \F_q^m}\omega^{\Tr(x,y)}\ket{y},$$
where $(x,y)$ stands for the standard scalar product
$x^Ty=\sum_{i=1}^m x_iy_i$ on $\F^m$. Our input superposition
(\ref{eq:cosetsup})
gets mapped to the state
$$\sum_{y\in \F^m}c_y\ket{y},$$
where
$$c_y=
\frac{\omega^{(v_0,y)}}{\sqrt{\size{{W}}q^m}}
\sum_{v\in {W}}\omega^{(v,y)}.
$$
The subspace ${W}^\perp$ consisting
of the vectors $u$ from $\F_q^m$ such that
$(u,v)=0$ for every $v\in W$ has dimension
$m-\dim_{\F_q}{W}$, therefore
$\size{{W}^\perp}=\frac{q^m}{\size{{W}}}$.
For $y\in {W}^\perp$ we have
$$|c_y|=
\frac{1}{\sqrt{\size{{W}}q^m}}
\sum_{v\in {W}}\omega^0=\frac{\size{{W}}}{\sqrt{\size{{W}}q^m}}=
\frac{1}{\sqrt{\size{{W}^\perp}}}.$$
It follows that
$$\sum_{y\in {W}^\perp}|c_y|^2=\size{{W}^\perp}\cdot
\frac{\size{1}}{{\size{{W}^\perp}}}=1.$$
Therefore for $y\not\in {W}^\perp$ we have
$c_y=0$ and if we measure $\ket{y}$,
we obtain a uniformly random element 
of ${W}^\perp$.

Assume now that ${\cal W}$ is a subspace of the linear
space $\Mat_{n\times n}(\F_q)$ of $n\times n$ matrices 
over $\F_q$. We can consider $n\times n$ matrices
as vectors of length $n^2$. Then the standard
scalar product of two matrices $A=(a_{ij})$
and $B=(b_{ij})$ is 
$$\sum_{i,j=1}^na_{ij}b_{ij}=
\tr(AB^T).$$
Here, for a matrix $D\in \Mat_{n\times n}(\F_q)$,
by $\tr(D)$ we denote the sum of the diagonal
elements of $D$. (Thus $\tr(D)$ is an element
of $\F_q$. The map $\tr$ from $\Mat_{n\times n}(\F_q)$ to
$\F_q$ should not be confused with the 
trace map $\Tr$ from $\F_q$ to $\F_p$, although
they are not completely unrelated.) We will make use
of the identity $\Tr(XY)=\Tr(YX)$.

\section{Finding hidden Borel subgroups in the general linear group}
\label{sect:Borel-GL}

In this section we outline a quantum algorithm for finding
a hidden Borel subgroup $\cal H$ in the group $\GL_n(\F_q)$.
Like the most hidden subgroup algorithms, our procedure
is based on using superpositions over cosets of $\cal H$, 
that is, states of the form
$$\ket{{\cal H}B}=
\frac{1}{\sqrt {\size {\cal H}}}\sum_{A\in \cal H}\ket{AB},$$
where $B$ is a matrix from $\GL_n(\F_q)$.
We will think of such a superposition as an approximation
of a superposition over a linear space of matrices and
apply the quantum Fourier transform of the linear
space $\Mat_{n\times n}(\F_q)$ to obtain a guess for the
last subspace in the flag stabilized by $\cal H$. The guess
will be verified in a straightforward way. If the guess turns
out to be correct, the further members of the flag can be
obtained by recursion.

\subsection{Obtaining coset superpositions}

The standard approaches to the hidden subgroup problem
in a group $\cal G$ start with the the
state
$\frac{1}{\sqrt{\size{\cal G}}}\sum_{x\in {\cal G}}\ket{x}\ket{0},$
apply the oracle for the function $f$ to obtain
$\frac{1}{\sqrt{\size{\cal G}}}\sum_{x\in {\cal G}}\ket{x}\ket{f(x)},$
and finally measure the second register to obtain
the coset superposition
$$\frac{1}{\sqrt{\size{\cal H}}}\sum_{x\in {\cal H}}\ket{xy}$$
with some $y\in {\cal G}$ ($\cal H$ is the subgroup hidden by the function
$f$). If $\cal G$ is abelian then the uniform superposition
$\frac{1}{\sqrt{\size{\cal G}}}\sum_{x\in {\cal G}}\ket{x}$
over ${\cal G}$ can be obtained by
applying the quantum Fourier transform of $\cal G$
to $\ket{0}$ (here $0$ stands for the neutral element
of $\cal G$). There are efficient methods for computing
uniform superpositions of certain further classes of groups, 
e.g., the algorithm of Watrous \cite{watrous} for solvable groups. 

For the purposes of our algorithm  it will be sufficient to 
approximate the uniform superposition 
$\frac{1}{\sqrt{\size{\GL_n(\F_q)}}}\sum_{x\in\GL_n(\F_q)}\ket{x}$
over the group $\GL_n(\F_q)$ by the uniform superposition
$\frac{1}{\sqrt{\size{\Mat_{n\times n}(\F_q)}}}\sum_{x\in\Mat_{n\times
n}(\F_q)}\ket{x}$
over $\Mat_{n\times n}(\F_q)$,
which can be efficiently computed using the quantum Fourier transform
of $\F_q^{n^2}$. The fidelity between the two states
is 
$$\sqrt{\frac{\size{\GL_n(\F_q)}}{\size{\Mat_{n\times n}(\F_q)}}}=
\sqrt{\frac{\prod_{i=0}^{n-1}(q^n-q^{i})}{q^{n^2}}}=
\sqrt{\prod_{j=1}^n(1-q^{-j})}>
\sqrt{\prod_{j=1}^\infty(1-2^{-j})}>\frac{1}{2}.$$
Therefore, if we first apply the quantum oracle $U_f$ to
the superposition
\break
$\frac{1}{\sqrt{\size{\Mat_{n\times n}(\F_q)}}}\sum_{x\in\Mat_{n\times
n}(\F_q)}\ket{x}\ket{0}$
and then measure the second register,
 we obtain a superposition over a coset of the hidden subgroup
$\cal H$ in $\GL_n(\F_q)$ with probability at least
$\frac{1}{4}$.

\subsection{Guessing the last subspace in the flag}
\label{subsect:guess}

Recall that our assumption is that
there exists an $n\times n$ invertible
matrix $X$ such that ${\cal H}=X^{-1}{\cal L}X$,
where 
$${\cal L}=\left\{A=(a_{ij})\in \GL_n(\F_q):
a_{ij}=0\mbox{~when~}i<j\right\}.$$
Then the last nontrivial member $U_{n-1}$ of the flag
$\F_q^n>U_1>\ldots>U_{n-1}>(0)$ stabilized by $\cal H$ 
is $X^{-1}V_{n-1}$, where $V_{n-1}$ consists
of the column vectors from $\F_q^n$
whose first $n-1$ entries are zero.

We will consider the multiplicative group
$\cal L$ as an approximation of the subspace
$${\cal L'}=\left\{A=(a_{ij})\in \Mat_{n\times n}(\F_q):
a_{ij}=0\mbox{~when~}i<j\right\}$$
of lower triangular matrices. Then 
${\cal H}$ will be thought of as an approximation
of ${\cal H'}=X^{-1}{\cal L}X$.
We have ${\size{\cal H}}={\size{\cal L}}=(q-1)^nq^{n(n-1)/2}$
and ${\size{{\cal H}'}}={\size{{\cal L}'}}=q^{n(n+1)/2}$.

Accordingly, for every $B\in \GL_n(\F_q)$,
the coset superposition
$$\ket{{\cal H}B}=
\frac{1}{\sqrt{\size{\cal L}}}\sum_{A\in\cal L}\ket{X^{-1}AXB}$$
will be considered as an approximation of
$$\ket{{\cal H}'B}=
\frac{1}{\sqrt{\size{\cal L}'}}\sum_{A\in{\cal L}'}\ket{X^{-1}AXB}.$$
The fidelity between 
$\ket{{\cal H}B}$ and $\ket{{\cal H}'B}$ is
$$\frac{\sqrt{\size{{\cal H}B}}}{\sqrt{\size{{\cal H}'B}}}=
\frac{\sqrt{\size{{\cal H}}}}{\sqrt{\size{{\cal H}'}}}=
\left(\frac{q-1}{q}\right)^{\frac{n}{2}}.$$

Therefore, if we apply the quantum Fourier transform of $\Mat_{n\times n}(\F_q)$
discussed in the previous section to the coset superposition
$\ket{{\cal H}B}$, and do the measurement then, with
a chance at least $\Omega\left(\left(\frac{q-1}{q}\right)^{n}\right)$,
the result will be a uniformly random element of the subspace
$({\cal H}'B)^\perp$, as it would be the case when we started
with the state $\ket{{\cal H}'B}$.

Let $Y$ be a matrix from $\Mat_{n\times n}(\F_q)$. Then $Y\in ({\cal H}'B)^\perp=
(X^{-1}{\cal L}'XB)^\perp$
if and only if $\tr(X^{-1}AXBY^T)=0$ for every $A\in {\cal L}'$.
As 
$$\tr\left(X^{-1}AXBY^T\right)=
\tr\left(AXBY^TX^{-1}\right)=
\tr\left(A\left((X^T)^{-1}YB^TX^T\right)^T\right),$$
we obtain that $Y\in ({\cal H}'B)^\perp$
if and only if $(X^T)^{-1}YB^TX^T\in {{\cal L}'}^{\perp}$. Furthermore,
as multiplying matrices by $B^T$ and conjugating matrices by
$X^T$ are bijections, we can conclude that if $Y$
is a uniformly random element of $({\cal H}'B)^\perp$
then $(X^T)^{-1}YBX^T$ is a uniformly random element of 
${{\cal L}'}^{\perp}$. Observe that the elements of 
${{\cal L}'}^{\perp}$ are just
the strictly upper triangular $n\times n$ matrices.
A strictly upper triangular matrix $Z$ has rank
$n-1$ if and only if all the entries of $Z$
just above the principal diagonal
are nonzero. For a uniformly random
strictly upper triangular matrix 
this happens with probability 
$\left(\frac{q-1}{q}\right)^{n-1}>
\left(\frac{q-1}{q}\right)^{n}$.

Observe that if $Z$ is a strictly upper
triangular matrix of rank  $n-1$ then
the kernel of $Z^T$ is the set $V_{n-1}$ of column 
vectors from $\F^n$ whose first $n-1$ entries are
zero. 
Obviously, the matrix $(X^T)^{-1}YB^TX^T$ has the same
rank as $Y$. If the rank is $n-1$, then 
the kernel of $XBY^TX^{-1}=\left((X^T)^{-1}YBX^T\right)^T$ is $V_{n-1}$,
whence the kernel of $Y^T$ is the 1-dimensional subspace 
$X^{-1}V_{n-1}$, which is the last subspace of the flag stabilized by
${\cal H}$. 

In summary, by applying the quantum Fourier transform
to the coset state $\ket{{\cal H}B}$ and then measuring the
result, with probability $\Omega\left((1-q^{-1})^{2n}\right)$
we obtain a matrix $Y$ of rank $n-1$ with kernel
$U_{n-1}$.

\subsection{Putting things together}

In this part we show how to check and use
a guess for the last subspace
$U_{n-1}$ of the flag stabilized by the hidden Borel
subgroup ${\cal H}$ provided by the algorithm described
in the previous subsection.

As $U_{n-1}$ is one-dimensional, we
assume that the guess is given by a nonzero column vector $u$.
Let $Z$ be a matrix from $\GL_n(\F_q)$ whose last column
is $u$. Then $U_{n-1}=ZV_{n-1}$. We replace the hiding
function $f$ by $f'$ defined as $f'(A)=f(ZAZ^{-1})$. An oracle
for $f'$ can be obtained from the oracle for $f$ in an
obvious way using this definition. The subgroup hidden by $f'$ is $Z^{-1}{\cal H}Z$
and the last subspace of the flag stabilized by 
$Z^{-1}{\cal H}Z$ is $Z^{-1}U_{n-1}$. 
The guess for $U_{n-1}$ is correct if and only if $Z^{-1}U_{n-1}=
V_{n-1}$, that is, the subgroup $Z^{-1}{\cal H}Z$ hidden
by $f'$ is contained in the subgroup of matrices of the form
$$
\left(
\begin{array}{cc}
\mbox{\Large $A$}
& 0 \\
v & \alpha
\end{array}\right),
$$
where $A\in \GL_{n-1}(\F_q)$, $\alpha\in\F_q\setminus \{0\}$
and $v$ is row vector of length $n-1$. Testing correctness
of the guess can be carried out by calling the oracle
for the identity matrix and for the $n-1$ matrices of the form
$$
\left(
\begin{array}{cc}
\mbox{\Large $I$}
& 0 \\
v & 1
\end{array}\right)
$$
with $v=(1,0,0,\ldots,0)$, $(0,1,0,\ldots,0)$, $\ldots$,
$(0,0,\ldots,0,1)$. 
Note that we can obtain a correct guess with
expected $O\left( (1-q^{-1})^{-n}\right)$
repetitions of procedure described in the
previous subsection.

Assume that the guess is correct. Then we consider the subgroup
$\cal G$ of $GL_{n}(\F)$ consisting of the matrices
of the form
$$
\left(
\begin{array}{cc}
\mbox{\Large $A$}
& 0 \\
0 & 1
\end{array}\right),
$$
with $A\in\GL_{n-1}(\F_q)$. Taking the upper left $n-1$ by $n-1$
block of matrices gives an isomorphism between $\cal G$
and $GL_{n-1}(\F_q)$. Furthermore, ${\cal G}\cap Z^{-1}{\cal H}Z$ 
is a Borel subgroup of $\cal G$. The subspaces in the flag stabilized by 
${\cal G}\cap Z^{-1}{\cal H}Z$ are intersections of
those for $Z^{-1}{\cal H}Z$ 
with the subspace of the column vectors with zero as last entry. 
We determine this flag by recursion. Then we obtain
the flag for $Z^{-1}{\cal H}Z$ by adding $U_{n-1}$ to each of the members.
Finally the complete flag for $\cal H$ is obtained by applying $Z$.

The group $\PGL_n(\F_q)$ is the factor of $\GL_n(\F_q)$
by its center consisting of the scalar matrices and the Borel subgroups
of $\PGL_n(\F_q)$ are just the images of the
Borel subgroups of $\GL_n(\F_q)$ under the quotient map.
As the scalar matrices from $\GL_n(\F_q)$
are contained in every Borel subgroup, the hidden
Borel subgroup problem for the groups $\PGL_n(\F_q)$ and 
$\GL_n(\F_q)$ essentially coincide. (A function hiding
a Borel subgroup of $\PGL_n(\F_q)$ can be lifted to
$\GL_n(\F_q)$ in the straightforward way.)
We have proved the following.

\begin{theorem}
The hidden Borel subgroup problem in $\GL_n(\F_q)$ 
(and in $\PGL_n(\F_q)$) can
be solved in quantum time $\poly(n+\log q+(1-q^{-1})^{-n})$.
\end{theorem}

\section{Finding Borel subgroups in the special linear group}
\label{sect:Borel-SL}

The special linear group $\SL_n(\F_q)$ consists
of $n$ by $n$ matrices over $\F_q$ with
determinant one. In this section we briefly
outline an extension of our method to finding
hidden Borel subgroups in $\SL_n(\F_q)$. A Borel
subgroup of $\SL_n(\F_q)$ is just the intersection
of $\SL_n(\F_q)$ with a Borel subgroup of $\GL_n(\F_q)$,
that is, the stabilizer of a flag 
$\F_q^n>U_1>\ldots>U_{n-1}>(0)$ of subspaces
within $\SL_n(\F_q)$. Again, we require the output
of the HSP algorithm to be this flag.

Assume that we have a function
$f$ defined on $\SL_n(\F_q)$ that hides a conjugate of
the subgroup ${\cal L}_0$
consisting of the lower triangular matrices
having determinant $1$. Let $\cal G$ stand
for the subgroup of $\GL_n(\F_q)$ consisting of matrices
whose determinant are from ${{\F_q^*}^n}$, where
${\F_q^*}^n=\{x^n:0\neq x\in\F_q\}$. We extend
$f$ to $\cal G$ as follows. Let $A$ be a matrix
from $\cal G$. We compute $\det A$ and find
en element $z\in \F_q$ such that $z^n=\det A$.
Such elements $z$ can be found e.g., by Berlekamp's
polynomial factoring algorithm~\cite{ber68,ber70}. We put $f(A)=f(z^{-1}A)$.
It turns out that the definition of $f(A)$ does not depend
on the choice of $z$. Indeed, if $z_1^n=z^n$ then
$z_1^{-1}zI$ is in the subgroup of $\SL_n(\F_q)$ hidden
by $f$ and therefore $f(z_1^{-1}A)=f(zA)$. The subgroup
of $\cal G$ hidden by the extended function will
be a conjugate of the lower triangular matrices
with determinant from ${\F_q^*}^n$. The fidelity
between the uniform superposition over this set
and the uniform superposition over all the lower
triangular matrices is at least $\frac{1}{\sqrt
n}\left(\frac{q-1}{q}\right)^n$.
Therefore, if we apply the method of Subsection~\ref{subsect:guess}
for guessing the last element of the stabilized flag,
we have a further factor $\Omega\left(\frac{1}{n}\right)$ for
the probability of obtaining a correct guess. Testing correctness 
and the recursion
are also essentially the same as in the case for $\GL_n(\F_q)$. 
We obtain the following.

\begin{theorem}
The hidden Borel subgroup problem in $\SL_n(\F_q)$ 
(and in $\PSL_n(\F_q)$) can
be solved in quantum time $\poly(n+\log q+(1-q^{-1})^{-n})$.
\end{theorem}

As $(1-q^{-1})^{-n}$ is polynomial in $n$ if
$q=\Omega(\frac{n}{\log n})$, we have

\begin{corollary}
When $q=\Omega(\frac{n}{\log n})$, the hidden Borel subgroup 
problem in $\GL_n(\F_q)$ and $\SL_n(\F_q)$ (and in 
$\PGL_n(\F_q)$ and $\PSL_n(\F_q)$) can
be solved in quantum time $\poly(n+\log q)$. In particular,
for constant $n$, 
the quantum complexity 
of the 
problem 
is $\poly(\log q)$. 
\end{corollary}

\section{Concluding remarks}
\label{sect:concl}

In this paper we have proved that the hidden Borel subgroup
in $\GL_n(\F_q)$ and $\SL_n(\F_q)$ can be solved in quantum
polynomial time if the size $q$ of the base field is not
too much smaller than the degree $n$. Perhaps the most important
question which if left open is existence of
polynomial time algorithms over small base fields 
(e.g., over fields of constant size). 

Other interesting questions are whether it is possible to extend
the result to the hidden Borel subgroup problem in other classical
groups (e.g., the orthogonal groups) and if it is possible to
find efficiently hidden conjugates of certain subgroups of
the lower triangular matrices such as the unitriangular matrices or
the diagonal matrices.  



\end{document}